\documentclass[pre,a4paper,twocolumn,superscriptaddress,%
floatfix]{revtex4}

\usepackage[T1]{fontenc}
\usepackage[utf8]{inputenc}

\usepackage{graphicx}
\usepackage{color}
\usepackage{amsmath}

\begin{document}

\title{Fluctuations of thermal variables investigated by cross-correlation function}

\author{Jean-Luc Garden}
\email{jean-luc.garden@neel.cnrs.fr}
\affiliation{Univ. Grenoble Alpes, CNRS, Grenoble INP, Institut N\'{E}EL,
38000 Grenoble, France}

\date{\today}

\begin{abstract}
Fluctuations in conjugate thermodynamic variables are studied using the cross-correlation function. A new procedure is given enabling the derivation of fluctuation formulas for a system in equilibrium. Specifically, the cross-correlation function between heat and temperature is employed for thermal variables. Additionally, fluctuation-dissipation relations involving the frequency-dependent specific heat are established. Moreover, a general relation concerning the average entropy production is also given, which is the microscopic analogue of the dissipation formula of the linear response theory. In the case of thermal variables, this formula finds application in various scenarios describing fluctuating thermal systems in equilibrium.
\end{abstract}

\maketitle

\section{Introduction}
The macroscopic state variables of a thermodynamic system fluctuate. Highly sensitive instruments can have access to these fluctuations. They impose a natural limit to the smallest signals accessible by the instruments. They are due to the corpuscular nature of particles at the microscopic scale which are submitted to thermal agitation. From seminal works in the field of statistical physics, particularly those addressing Brownian motion \cite{ORNSTEIN}, or those on the thermal agitation of electric charges within a conductor \cite{JOHNSON,NYQUIST}, it is well-established that these fluctuations are intrinsically linked to damping parameters. For example, the thermal conductance $K$ for thermal systems, or the resistance $R$ in electrical circuits are such damping parameters. For a system in equilibrium, fluctuations manifest, on average, as secondary order terms in the deviation from the mean value. If we call $\xi$ an extensive thermodynamic variable and $F$ the associated conjugate intensive variable, fluctuations are represented as $\delta\xi=\xi-\overline{\xi}$ and $\delta F=F-\overline{F}$, where the over-line denotes a temporal averaging over extended durations (similar to $\left\langle  \right\rangle$  ensemble averaging over all the possible fluctuations following ergodic hypothesis).
 
In 1952, Callen and Greene formulated relations connecting fluctuations with dissipative parameters for all sets of conjugate thermodynamic variables \cite{CALLEN2}: 

\begin{subequations}
\begin{equation}
  \label{eq:CALLENKSI}
  \overline{\left(\delta\xi\right)^{2}}=\frac{-2k_{B}}{\pi}\displaystyle \int d\omega\sigma_{s}(\omega)\omega^{-2} ,
\end{equation}
\begin{equation}
  \label{eq:CALLENF}
  \overline{\left(\delta F\right)^{2}}=\frac{-2k_{B}}{\pi}\displaystyle \int d\omega R_{s}(\omega) .
\end{equation}
\end{subequations}
$\xi$ is the extensive variable and $F$ is the hypothetical random force acting on the dissipative system. $k_{B}$ is the Boltzmann constant and $\omega=2\pi f$ is the angular frequency. The conductance $\sigma_{s}$ is the real part of the generalized admittance, and $R_{s}$ is the real part of the generalized impedance when canonical constraint on $\xi$ is considered \cite{CALLEN2}. These relations~(\ref{eq:CALLENKSI}) and~(\ref{eq:CALLENF}) are a thermodynamic formulation of the fluctuation-dissipation theorem obtained by Callen and Welton from quantum statistical physics \cite{CALLEN3}. The thermodynamic approach was then generalized to several extensive variables fluctuations \cite{CALLEN4}. 

In the present paper, we shall use the same formalism based on thermodynamics in order to propose an new procedure starting from fluctuations of random variables, giving the frequency dependence of their spectral densities, and leading then to their mean square values. Contrary to the usual approach, we will not start from the autocorrelation functions of each random variable, but from their cross-correlation function. Moreover, the equipartition theorem will not be used. Instead, we will use an equivalent expression based on the statistical expectation value of the entropy decrease due to the contribution of all the possible fluctuations. Although the paper focuses on thermal variable fluctuations, the generality of the approach proposed here will be shown across the investigation of fluctuation formulas for a simple $RC$ electrical circuit in appendix A. In the specific case of thermal variables, the procedure may apply on a wide class of thermal processes. This point is illustrated in appendix C across the investigation of the Cattaneo-Vernotte process of propagation of heat in a medium. 

\section{Objectives and main results of the paper}
\label{sec:Objective}
The origin of our work has resulted from the observation that in Eq.~(\ref{eq:CALLENKSI}) and~(\ref{eq:CALLENF}), the inherent generalized impedances (or admittances) differ from each other. Notwithstanding the conjugate nature of $\xi$ and $F$, the generalized impedances do not characterize the same process. This particular aspect remains unaddressed in the reference \cite{CALLEN2}. Therefore, in a first step we shall propose a new procedure which takes into account this observation in order to obtain new general fluctuation formulas. The second objective of the paper will be to apply specifically these formulas on thermal variable fluctuations. Our procedure will be used in order to obtain well-known formulas for the fluctuations of energy $\delta E=E-\overline{E}$ and temperature $\delta T=T-\overline{T}$ of a thermodynamic system connected to a thermostat. The third objective of this study will be to establish a precise microscopic foundation for the frequency-dependent specific heat, denoted as $C(\omega )$. Three new fluctuation-dissipation relations will be obtained using this generalized thermal susceptibility. Eventually, the last objective of the paper will be to give from our procedure three new formulas involving entropy production. The first one is general, while the two others belong to thermal variables.  All these results are summarized below and their derivations are given in section III.

\subsection*{Fluctuation formulas} 

The analogue of formulas~(\ref{eq:CALLENKSI}) and~(\ref{eq:CALLENF}) derived from our procedure are:

\begin{subequations}
\begin{equation}
  \label{eq:GARDENKSI}
  \overline{\left(\delta\xi\right)^{2}}=\frac{1}{2\pi}\displaystyle \int_{0}^{+\infty} d\omega S_{FF}(\omega)\left|Y_{s}(\omega)\right|^{2}\omega^{-2} ,
\end{equation}
\begin{equation}
  \label{eq:GARDENF}
  \overline{\left(\delta F\right)^{2}}=\frac{1}{2\pi}\displaystyle \int_{0}^{+\infty} d\omega\frac{S_{\dot{\xi}\dot{\xi}}(\omega)}{\left|Y_{p}(\omega)\right|^{2}} .
\end{equation}
\end{subequations}
$Y_{s}(\omega)$ is the admittance of a "series-system", while $Y_{p}(\omega)$ is that of a "parallel-system". The definition of such types of systems will be provided in the upcoming sections. $S_{FF}$ and $S_{\dot{\xi}\dot{\xi}}$ are the spectral densities of the force and flux respectively. The flux, $\dot{\xi}=d \xi/dt$, is the time derivative of the extensive variable. The $\omega$-dependency of the force or flux spectral densities is just a matter of physical situation as we will see later. The microscopic nature of matter does not appear in those previous formulas, while it appears through the constant $k_{B}$ in the formulas~(\ref{eq:CALLENKSI}) and~(\ref{eq:CALLENF}) of Callen and Greene. This is why a third formula must be added in order that our procedure be complete:

\begin{equation}
  \label{eq:GARDENKB}
  \frac{1}{2\pi}\displaystyle \int_{0}^{+\infty} d\omega\frac{S_{\dot{\xi}\dot{\xi}}(\omega)}{\omega}\frac{\Im\{Y_{p}(\omega)\}}{|Y_{p}(\omega)|^{2}}=-k_{B} .
\end{equation}
$\Im\{Y_{p}(\omega)\}$ is the imaginary part of the complex function $Y_{p}(\omega)$. This formula is obtained from the cross-correlation function of $\xi$ and $F$. Like autocorrelation functions, cross-correlation functions and cross-spectral densities are related by means of the Wiener-Khintchine relations \cite{MAX}. The $\omega$-dependency of the flux of the extensive variable is again a matter of choice, depending on the considered physical case.

\subsection*{Thermal variable fluctuations}

For a system of heat capacity $C$, thermally coupled to a thermal bath, energy and temperature undergo fluctuations (cf.~Fig1). Their mean-square values are:

\begin{subequations}
\begin{equation}
  \label{eq:EFLUC}
  \overline{\left(\delta E\right)^{2}}=k_{B}T^{2}C ,
\end{equation}
\begin{equation}
  \label{eq:TFLUC}
  \overline{\left(\delta T\right)^{2}}=\frac{k_{B}T^{2}}{C} .
\end{equation}
\end{subequations}

 Temperature fluctuations arise from the incessant random exchanges of heat carriers \cite{FOOTNOTE1} between the system and its thermal reservoir. To understand temperature fluctuations, it is sometimes useful to introduce a fictitious random power $\delta P(t)$ (cf.~Fig1), defined in such way that fluctuations can be viewed as the outcomes of this stochastic "force" perturbing the system, similarly to Langevin's random force in the interpretation of Brownian motion \cite{LANGEVIN}. The fluctuation formula concerning the mean square value of this random power is written:

\begin{equation}
  \label{eq:PFLUC}
  \overline{\left(\delta P\right)^{2}}=4k_{B}T^{2}K\Delta f .
\end{equation}
where $\Delta f$ is a frequency bandwidth on the power spectrum and $K$ the thermal coupling coefficient (damping parameter) between the system and its bath (cf.~Fig1). Formulas~(\ref{eq:EFLUC}) and~(\ref{eq:TFLUC}) can be directly obtained from the classical thermodynamic fluctuation theory based on canonical distributions, but not the formula (\ref{eq:PFLUC}) \cite{CALLEN1, LANDAU, CALLEN5, MISHIN}. Formulas~(\ref{eq:EFLUC}) and~(\ref{eq:TFLUC}) are mean square values of the total fluctuations of energy and temperature, and they tell us nothing about their frequency distribution on the spectrum. This is precisely the route from microscopic agitation to these final formulas we would like to trace in this paper. In particular from our procedure we shall obtain formulas~(\ref{eq:EFLUC}),~(\ref{eq:TFLUC}), and~(\ref{eq:PFLUC}) for thermal variables without using the equipartition theorem as it is usually made. The equipartition theorem, indeed, represents an alternative manifestation of the fluctuation formulas we aim to derive from a microscopic basis. Notably, Eq.~(\ref{eq:TFLUC}) corresponds to the equipartition theorem governing temperature fluctuations. Our procedure based on cross-correlation functions provides the spectral densities of each fluctuating variables.

\subsection*{Frequency dependent specific heat}

The frequency dependent complex specific heat is a quantity measured from modulated temperature calorimetry, or ac-calorimetry \cite{GARDEN1,GARDEN4,SULLIVAN,LION,DEOLIVEIRA}. In these types of calorimetry experiments, the fictitious power $\delta P$ is a real power oscillating at a frequency $\omega$, and the oscillating temperature of a sample is recorded with a thermometer \cite{SULLIVAN}. Here we show that $C(\omega)$ is a type of susceptibility, as such, it can be expressed from the admittance and in particular, from Eqs.~(\ref{eq:GARDENKSI}) and~(\ref{eq:GARDENF}), we have:

\begin{subequations}
\begin{equation}
  \label{eq:C1}
  \overline{\left(\delta E\right)^{2}}=\frac{1}{2\pi}\displaystyle \int_{0}^{+\infty} d\omega S_{TT}(\omega)\left|C_{s}(\omega)\right|^{2} ,
\end{equation}
\begin{equation}
  \label{eq:C2}
  \overline{\left(\delta T\right)^{2}}=\frac{1}{2\pi}\displaystyle \int_{0}^{+\infty} d\omega S_{EE}(\omega)\left|C_{p}(\omega)\right|^{-2} .
\end{equation}
\end{subequations}

From Eq.~(\ref{eq:GARDENKB}), we have:

\begin{equation}
  \label{eq:C3}
  \frac{1}{2\pi}\displaystyle \int_{0}^{+\infty} d\omega S_{TT}(\omega)C'_{p}(\omega)=k_{B}T^{2} .
\end{equation}

These three relations constitute fluctuation-dissipation theorems for thermal variables \cite{NIELSEN} (cf.~Section Discussion). $S_{TT}(\omega)$ is the temperature spectral density, $S_{EE}(\omega)$ is the energy spectral density, and $C'_{p}(\omega)$ is the real part of the complex specific heat of a parallel-system, while $\left|C_{p}(\omega)\right|^{2}$ and $\left|C_{s}(\omega)\right|^{2}$ are the square modulus of the complex specific heat of a parallel and series-system respectively.

\subsection*{Entropy production}

As we have used cross-correlation function between $\delta\xi$ and $\delta F$ in order to derive Eqs.~(\ref{eq:GARDENKSI}), (\ref{eq:GARDENF}) and~(\ref{eq:GARDENKB}), then we can use cross-correlation function between $\delta\dot{\xi}$ and $\delta F$ in order to derive a formula for the entropy produced by an equilibrium system, on average over a long time:

\begin{equation}
  \label{eq:ENTROP1}
  \overline{\sigma_{i}}=-\frac{1}{4\pi}\displaystyle \int_{0}^{+\infty} d\omega S_{\dot{\xi}\dot{\xi}}(\omega)\Re\{Z_{p}(\omega)\} ,
\end{equation}
where $\Re\{Z_{p}(\omega)\}$ is the real part of the impedance belonging to a parallel-system.

The direct application of this formula to thermal variables yields:

\begin{equation}
  \label{eq:ENTROP2}
  \overline{\sigma_{i}}=\frac{1}{4\pi T^{2}}\displaystyle \int_{0}^{+\infty} d\omega S_{TT}(\omega)\omega C"_{p}(\omega) ,
\end{equation}
where $C"_{p}(\omega)$ is the imaginary part of the specific heat belonging to a parallel-system.

The last formula of the paper results from a time-average around one particular frequency $\omega_{0}$ on the spectrum. It is obtained in limiting the integration in Eq.~(\ref{eq:ENTROP2}) to one period $T_{e}=\frac{2\pi}{\omega_{0}}$:

\begin{equation}
  \label{eq:ENTROP}
  \Delta_{i}S_{\omega=\omega_{0}}=\pi\frac{S_{TT}(\omega_{0})}{T^{2}}C"_{p}(\omega_{0}) .
\end{equation}

It provides the net positive amount of entropy per unit of frequency generated by temperature fluctuations at $\omega_{0}$ on the spectrum. It is analogous to an expression obtained in temperature modulated experiments \cite{GARDEN1}. Therefore, it is of a certain degree of generality because in the case of temperature modulated experiments it applies to a wide class of thermal processes \cite{GARDEN2,GARDEN3,GARDEN4}.

\section{Fluctuation formulas obtained by cross-correlation function}
\label{sec:Thermal}

In order to derive formulas~(\ref{eq:GARDENKSI}),~(\ref{eq:GARDENF}) and~(\ref{eq:GARDENKB}),  we need to apply several assumptions:

\begin{itemize}
    \item the process of random variable fluctuations is stationary. The time average of the fluctuations is much longer than the macroscopic relaxation time of the system. Under these circumstances, time-averaging is equivalent to statistical-averaging over all the possible fluctuations (ergodic hypothesis). The expectation value does not depend on an initial value. In other word, thermodynamic equilibrium holds and remains in such a state indefinitely   
    \item the random variable fluctuations obey a statistical Gaussian distribution. This is a consequence of the huge numbers of chocs at the microscopic scale
    \item the random thermodynamic variables are real analytical time dependent signals
    \item the process of fluctuations is linear. The admittance, impedance or susceptibility are issued from the linear macroscopic response of the system to the perturbing force. This is the case for $C(\omega)$ 
    \item the spectral densities are all constant at low frequencies (white noise assumption). This holds if we consider that correlation functions rapidly decrease along time. This happens if the numerous chocs of the particles at a molecular level, are so rapid, that under the time average of the fluctuations there is no correlation anymore (for one or several variables)
    \item for the investigation of thermal variables, no work fluctuations are considered. Under these circumstances, heat fluctuations between the system and its bath are equivalent to energy fluctuations. Conversely, when other variables are considered (for example charge fluctuations in appendix A), work fluctuations is equivalent to energy fluctuations. However, in the two cases the average entropy production leads to the same energy dissipated of $k_{B}T/2$ per degree of freedom 
\end{itemize}

\subsection*{Probability of fluctuations}

When the state variables of a dissipative system undergo fluctuations around equilibrium, each fluctuation results in a reduction of entropy. Over time, through the process of averaging, the system dissipates heat (via the dissipative parameters), leading to the production of entropy. This phenomenon ensures the system's perpetual maintenance in an equilibrium state, characterized by constant expectation values of its state variables. Paradoxically, the existence of equilibrium relies on the ongoing occurrence of non-equilibrium processes. Thermodynamic equilibrium owes its sustainability to the continuous dissipation of heat from the system to the thermal bath. Landau and Lifshitz quantitatively assess this entropy production by considering the maximum work that a body can transfer to an external medium or, equivalently, the minimum work that an external source must provide to the body \cite{LANDAU}. Kondepudi and Prigogine, on the other hand, opt for directly addressing entropy production \cite{PRIGO}. Landau and Lifshitz employ a hypothetical reversible process to address thermodynamic fluctuations, whereas Prigogine and Kondepudi directly consider the non-equilibrium path.  This latter approach is employed in this paper. The entropy decrease due to the fluctuation of an extensive variable $\delta x$ associated with its conjugate intensive variable $\delta y$ is written $\Delta_{i}S$. The density probability of such a fluctuation is written \cite{PRIGO}:

\begin{equation}
  \label{eq:PROBA}
  p=Ae^{\frac{\Delta_{i}S}{k_{B}}} .
\end{equation}

Since the system is in equilibrium, the entropy decrease $\Delta_{i}S$ must be developed in series expansion up to a second order in the fluctuations of $\delta x$ and  $\delta y$:

\begin{equation}
  \label{eq:ENTROPYPROD1}
  \Delta_{i}S\sim\frac{1}{2}\delta^{2}S=\frac{1}{2}\delta x\delta y<0 .
\end{equation}

$\delta x$ and $\delta y$ are random variables obeying Gaussian distributions by assumption. The statistical average over all the possible values of the fluctuations $\delta x$ and  $\delta y$  of this entropy fall with the probability distribution~(\ref{eq:PROBA}) yields to:

\begin{equation}
  \label{eq:ENTROPYPROD2}
  \left\langle \Delta_{i}S\right\rangle =-\frac{k_{B}}{2} .
\end{equation}

The calculation is given in appendix B.

\subsection*{Cross-correlation function and cross-spectral density}

Here we developed specifically our procedure for thermal variable fluctuations for simplicity reasons, but the general formulas~(\ref{eq:GARDENKSI}),~(\ref{eq:GARDENF}) and~(\ref{eq:GARDENKB}) can be obtained exactely from the same way replacing $E$ by $\xi$ and $1/T$ by $F$. Indeed, for thermal variables, the extensive variable is the energy $E$ with the conjugate intensive variable $1/T$. The cross-correlation function of the random variables $\delta E(t)$ and $\delta \frac{1}{T}(t)$ results in a time average of their product but they are time-shifted. The stationary condition mandates that the function relies solely on the temporal drift, while remaining independent of the initial instant of the average \cite{MAX}:

\begin{equation}
  \label{eq:INTERCO1}
  \psi_{E\frac{1}{T}}\left(\tau\right)=\lim_{{t' \to +\infty}}\frac{1}{t'} \displaystyle \int_{0}^{t'} dt \delta E(t)\delta \frac{1}{T}(t-\tau)
 .
\end{equation}

This function is generally a rapidly decreasing function of $\tau$. It gives an idea of how energy fluctuations are correlated to temperature fluctuations taken at different instants. Since in physics temperature is the relevant measured variable, let us work from now with the energy/temperature cross-correlation function remarking that:

\begin{equation}
  \label{eq:INTERCO2}
  \psi_{E\frac{1}{T}}\left(\tau\right)=-\frac{1}{T^{2}}\lim_{{t' \to +\infty}}\frac{1}{t'} \displaystyle \int_{0}^{t'} dt \delta E(t)\delta T(t-\tau)=-\frac{\psi_{ET}(\tau)}{T^{2}}
 .
\end{equation}

Here, and in the following, $T$, the absolute temperature, is the temperature of the bath, but we omit the index $b$ for the sake of simplification. For the specific value $\tau=0$, the limit of the integral in Eq.~(\ref{eq:INTERCO1}) is a time average over infinite time, which by assumption is equivalent to a statistical average of two conjugate variables. Owing to the result~(\ref{eq:ENTROPYPROD2}), we have:

\begin{equation}
  \label{eq:PSI1}
  \psi_{ET}(0)=k_{B}T^{2} .
\end{equation}

We arrive then to the result that for a time drift $\tau$ equal to zero, the cross-correlation function of energy and temperature is equal to $k_{B}T^{2}$.

From Parseval's relation and using the Wiener-Khinchin theorem, the cross-correlation function for the specific value $\tau=0$ can be written as follows \cite{MAX}:

\begin{equation}
  \label{eq:PSI2}
  \psi_{ET}(0)=\displaystyle \int_{-\infty}^{+\infty}d\omega\delta E(\omega)\delta T^{*}(\omega) ,
\end{equation}
where by definition the cross-spectral density between energy and temperature is \cite{MAX}:

\begin{equation}
  \label{eq:SPECTRAL1}
  S_{ET}(\omega)=\delta E(\omega)\delta T^{*}(\omega) .
\end{equation}

$\delta E(\omega)$ and $\delta T(\omega)$ are the Fourier transforms of the fluctuating variables $\delta E(t)$ and $\delta T(t)$:

\begin{subequations}
\begin{equation}
  \label{eq:TFE}
  \delta E(\omega)=\displaystyle \int_{-\infty}^{+\infty} dt \delta E(t)e^{-i\omega t} ,
\end{equation}
\begin{equation}
  \label{eq:TFT}
  \delta T(\omega)=\displaystyle \int_{-\infty}^{+\infty} dt \delta T(t)e^{-i\omega t} .
\end{equation}
\end{subequations}
The star indicates complex conjugation. Contrary to spectral densities derived from the autocorrelation function of single-variable which are real,  cross-spectral densities are complex functions. $S_{ET}(\omega)$ has real and imaginary parts. However, since $\delta E(t)$ and $\delta T(t)$ are real signals, the cross-spectral density benefits of the hermitic symmetry \cite{MAX,DEGROOT}. The real part of $S_{ET}(\omega)$ is an even function of the frequency, while the imaginary part of $S_{ET}(\omega)$ is an odd function of the frequency. Eq.~(\ref{eq:PSI2}) immediately simplifies to:

\begin{equation}
  \label{eq:PSI3}
  \psi_{ET}(0)=\displaystyle \int_{-\infty}^{+\infty}d\omega \Re\{S_{ET}(\omega)\} .
\end{equation}

Since negative frequencies have no meaning in physics, the cross-spectral density $S_{ET}(\omega)$ is replaced by a quantity which has two times the previous value for each positive frequency, and the integration is now taken only over positive frequencies:

\begin{equation}
  \label{eq:PSI4}
  \psi_{ET}(0)=\displaystyle \int_{0}^{+\infty}d\omega \Re\{S^+_{ET}(\omega)\} .
\end{equation}

This mathematical artifice yields precisely the same outcome. From now the sign $+$ is omitted as usual \cite{KITTEL1,REIF}.

\subsection*{Admittance, impedance and susceptibility}

Generally, the admittance is defined as the ratio of the Fourier transform of the flux of the extensive variable with the Fourier transform of the intensive force when the system responds to a macroscopic force:

\begin{equation}
  \label{eq:ADMIT}
  Y(\omega)=\frac{\delta\dot{\xi}(\omega)}{\delta F(\omega)} .
\end{equation}

It is however possible to define an admittance directly from the Fourier transforms of the random fluctuating signals. This is true and limited to a first order in a series expansion, which gives justification on the assumption of linear regime. This important point is discussed with details in \cite{CALLEN2}; it is a crucial point, as it enables the establishment of a connection between the fluctuations of variables at the microscopic scale and the system's response to an external force at the macroscopic scale. Hence, the admittance of fluctuating thermal variables is:

\begin{equation}
  \label{eq:ADMITT}
  Y(\omega)=-T^{2}\frac{\delta\dot{E}(\omega)}{\delta T(\omega)}=-i\omega T^{2}\frac{\delta E(\omega)}{\delta T(\omega)} ,
\end{equation}
where $\delta E(\omega)$ and $\delta T(\omega)$ are defined with Eqs.~(\ref{eq:TFE}) and~(\ref{eq:TFT}).
From these two previous Fourier transforms, a frequency dependent specific heat can be defined exactly like in a dynamic calorimetry experiment \cite{GARDEN1}, or linear response theory \cite{FOMINAYA}:

\begin{equation}
  \label{eq:COMEGA}
  C(\omega)=\frac{\delta E(\omega)}{\delta T(\omega)}=C'(\omega)-iC"(\omega) .
\end{equation}

From this definition, and knowing that the susceptibility, called $\chi(\omega)$, is defined as the ratio of the extensive and intensive variables respectively, we have:

\begin{equation}
  \label{eq:ADMITC}
  Y(\omega)=i\omega \chi(\omega)=-i\omega T^{2}C(\omega) .
\end{equation}

Thus $C(\omega)$ is, up to the factor $-T^{2}$, a generalized susceptibility \cite{GARDEN1}. The impedance is the inverse of the admittance even though sometimes it is defined as the ratio of the output on the input in noise measurements \cite{MACCOMBIE}.

\subsection*{Fluctuation formulas for thermal variables}

\begin{figure}[h]
  \centering
  \includegraphics[width=8.6 cm]{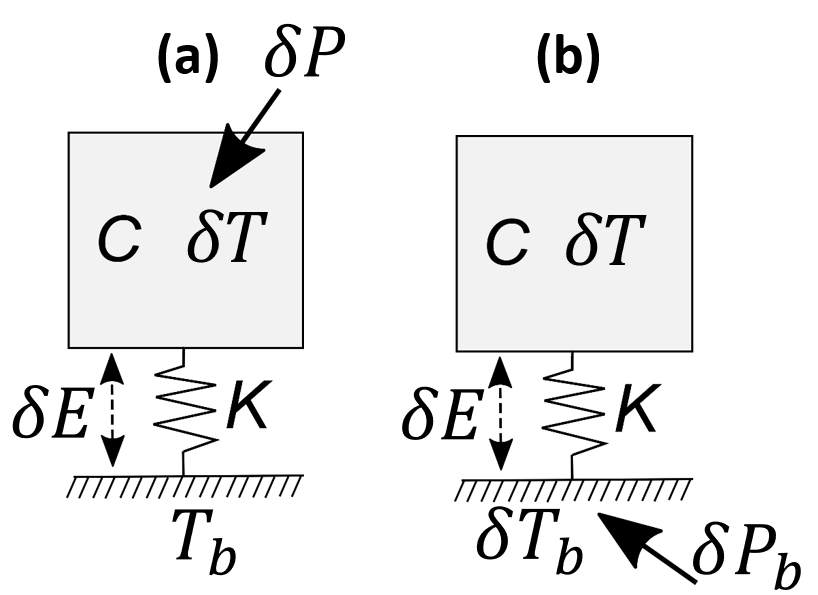}
  \caption{Picture of the dissipative system composed of a heat capacity $C$ coupled to a thermal bath of temperature $T_{b}$ by means of a heat exchange coefficient $K$. (a) Parallel-system. A fictitious thermal fluctuating power $\delta P$ separates in two contributions, one stored in $C$ and one as a heat loss to the bath across $K$. (b) Series-system. A fictitious temperature fluctuations of the bath induces energy (heat) fluctuations of the system. }
  \label{fig:Fig1}
\end{figure}

The Fig.~(\ref{fig:Fig1}) depicts two different situations leading to thermal variables fluctuations.
The Fig.~(1a) belongs to what we have called a parallel-system, while Fig.~(1b) belongs to a series-system. The difference between them is that the input "noisy power" is not supplied at the same location on the system/bath assembly. In the two situations the system is represented by a volume of specific heat $C$ and temperature $T$ coupled to a thermal bath of temperature $T_{b}$ by means of a thermal exchange coefficient $K$. For a parallel-system a fortuitous Langevin's force $\delta P$ is considered at the level of the system while for the series-system it is considered at the level of the bath. For a series-system the temperature fluctuations of the bath induce temperature fluctuations of the system and energy fluctuations. For a parallel-system this is the heat-flux $\delta P$ which induces temperature fluctuations of the system. For thermal variables the difference between the two types of systems is difficult to apprehend, but for the electrical case investigated in appendix A, it is easier to manage with parallel and series electrical connections of the resistor and capacitor.

\subsubsection*{Power and temperature fluctuations of a parallel-system}

Fig.~(1a) shows that temperature fluctuations $\delta T$ obey a first-order linear differential equation:

\begin{equation}
  \label{eq:EQUADIFF}
  C\frac{d(\delta T)}{dt}+K\delta T=\delta P .
\end{equation}

This is definitively a parallel system, as when a macroscopic heat flux is applied to the system, it splits into two components in parallel. A part of the heat flux is stored in the heat capacity (represented by the first term on the left-hand side of Eq.~(\ref{eq:EQUADIFF})), while the other part transfers to the thermal bath through the heat exchange coefficient $K$ (indicated as the second term on the left-hand side of Eq.~(\ref{eq:EQUADIFF})). However, Eq.~(\ref{eq:EQUADIFF}) applies for fluctuations around the mean value. By performing the Fourier transformation of this linear differential equation, it is straightforward to obtain:

\begin{subequations}
\begin{equation}
  \label{eq:CTHER1}
  C_{p}(\omega)=C-i\frac{K}{\omega} ,
\end{equation}
\begin{equation}
  \label{eq:ADMITTHER1}
  Y_{p}(\omega)=-T^{2}\left(i\omega C +K\right).
\end{equation}
\end{subequations}

The specific heat $C$ in the previous equations (see also Fig.~(\ref{fig:Fig1})) is the equilibrium specific heat of the dissipative system itself. The spectral density of the power $\delta P(t)$ is written $S_{PP}(\omega)$. Owing to the definition~(\ref{eq:SPECTRAL1}), and using the previous equation~(\ref{eq:ADMITT}), the cross-spectral density can be written as a function of the admittance:

\begin{equation}
  \label{eq:SPECTRAL2}
  S_{ET}(\omega)=T^{2}S_{PP}(\omega)\frac{iY_{p}(\omega)}{\omega\left|Y_{p}(\omega)\right|^{2}}
   .
\end{equation}

Taking the real part of the previous expression, and owing to Eqs.~(\ref{eq:PSI4}) and~(\ref{eq:PSI1}), we immediately arrive to:

\begin{equation}
  \label{eq:GARDENKBP}
  \frac{1}{2\pi}\displaystyle \int_{0}^{+\infty} d\omega\frac{S_{PP}(\omega)}{\omega}\frac{\Im\{Y_{p}(\omega)\}}{|Y_{p}(\omega)|^{2}}=-k_{B} ,
\end{equation}
which is the wanted Eq. (\ref{eq:GARDENKB}) with $S_{\dot{\xi}\dot{\xi}}(\omega)=S_{PP}(\omega)$ in the case of thermal variables. 

In Eq.~(\ref{eq:EQUADIFF}) it can be assumed that the spectral density of the fortuitous force $S_{PP}$ is constant, like in the classical treatment of the Langevin equation. This hypothesis is not valid for all thermal processes. In appendix C, we investigate the Cattaneo-Vernotte heat propagation process, for which we show that $S_{PP}$ is frequency dependent, leading to different fluctuation formulas. Now, upon exclusion of $S_{PP}$ from the previous integral, the subsequent step involves integrating the residual expression $\frac{\Im\{Y_{p}(\omega)\}}{\omega|Y_{p}(\omega)|^{2}}=-\frac{C}{(KT)^{2}(1+(\omega\tau)^{2})}$ utilizing Eq.~(\ref{eq:ADMITTHER1}). This leads to the derivation of the following expression for the power spectral density:

\begin{equation}
  \label{eq:SPP}
  S_{PP}=4k_{B}T^{2}K ,
\end{equation}
which by another integration on a bandwidth $\Delta f$ is the expected formula (\ref{eq:PFLUC}):

\begin{equation}
  \label{eq:PFLUCBIS}
  \overline{\left(\delta P\right)^{2}}=4k_{B}T^{2}K\Delta f . \nonumber
\end{equation}

Under these circumstances, it is trivial to obtain the spectral density of temperature fluctuations, which is linked to the power spectrum by:

\begin{equation}
  \label{eq:SPPSTT}
  S_{PP}=|Y_{p}(\omega)|^{2}S_{\frac{1}{T}\frac{1}{T}}=\frac{|Y_{p}(\omega)|^{2}}{T^{4}}S_{TT} .
\end{equation}

Since the power spectral density is constant, the frequency dependence of the temperature spectral density comes from the square modulus of the thermal admittance (using Eq.~(\ref{eq:ADMITTHER1})):

\begin{equation}
  \label{eq:STT}
  S_{TT}(\omega)=T^{4}\frac{S_{PP}}{|Y_{p}(\omega)|^{2}}=\frac{4k_{B}T^{2}K}{K^{2}+(\omega C)^{2}} .
\end{equation}

The integration of this expression between $0$ and $+\infty$ yields to:

\begin{equation}
  \label{eq:TFLUCBIS}
  \overline{\left(\delta T\right)^{2}}=\frac{k_{B}T^{2}}{C} , \nonumber
\end{equation}
which is the expected formula~(\ref{eq:TFLUC}).

Now, owing to the general relation~(\ref{eq:ADMITC}), formula~(\ref{eq:GARDENKBP}) transforms to:

\begin{equation}
  \label{eq:GARDENKBC1}
  \frac{1}{2\pi}\displaystyle \int_{0}^{+\infty} d\omega\frac{S_{PP}(\omega)}{\omega^{2}|C_{p}(\omega)|^{2}}C'_{p}(\omega)=k_{B}T^{2} .
\end{equation}

Let us notice with  Eq.~(\ref{eq:SPPSTT}) that:

\begin{equation}
  \label{eq:ACCALO}
  \frac{S_{PP}(\omega)}{\omega^{2}|C_{p}(\omega)|^{2}}=S_{TT}(\omega) .
\end{equation}

This leads directly to the desired Eq.~(\ref{eq:C3}):

\begin{equation}
  \label{eq:C3BIS}
  \frac{1}{2\pi}\displaystyle \int_{0}^{+\infty} d\omega S_{TT}(\omega)C'_{p}(\omega)=k_{B}T^{2} . \nonumber
\end{equation}

For the process described in Fig.~(\ref{fig:Fig1}) by the differential Eq.~(\ref{eq:EQUADIFF}), $C'_{p}=C$ independent of the frequency (see Eq.~(\ref{eq:CTHER1})). Excluding this part from the integral above and placing it in the denominator of the right-hand-side term directly yields the fluctuation formula~(\ref{eq:TFLUC}). Once the complex specific heat is determined, it sometimes becomes more straightforward to employ the latter equation instead of integrating a complex expression, such as  Eq.~(\ref{eq:STT}) to derive temperature fluctuations.  However, for other thermal processes, the real part of $C(\omega)$ can be frequency dependent. In appendix C, it is the case for the Cattaneo-Vernotte process of heat propagation.

At this step, let us make a small aside for calorimetry experimenters, remarking that Eq.~(\ref{eq:ACCALO}) above is analogous to the expression used in the measurement of specific heat in ac-calorimetry \cite{GARDEN1,SULLIVAN}, which we rewrite here for the sake of clarity:

\begin{equation}
  \label{eq:ACCALO2}
  \delta T_{ac}=\frac{P_{0}}{\omega |C(\omega)|}=\frac{P_{0}}{\omega C\sqrt{1+\frac{1}{(\omega\tau_{th})^{2}}}} ,
\end{equation}
with $\tau_{th}=C/K$ is the thermal relaxation time, and $P_{0}$ is the amplitude of the oscillating ac-power supplied on the sample. Generally experiments occur under the requirement $(\omega\tau_{th})^{2}>>1$ for adiabatic conditions \cite{SULLIVAN}. In Eq.~(\ref{eq:ACCALO}), however, only random noisy signals are considered, but not power and temperature macroscopic oscillations. 

\subsubsection*{Energy fluctuations of a series-system}

In order to obtain the spectral density of energy fluctuations, $S_{EE}(\omega)$, a transition from a parallel-system to a series-system is necessary (cf. Fig. (1b)). While in a parallel-system the flux is introduced at the body level where it splits into two parallel components, in contrast a series-system receives the flux directly from the bath, which dictates the system's temperature. In a series-system, the admittance is defined as the ratio of power into the bath to the inverse temperature of the bath (cf. Fig.(1b)). In this case, the temperature spectral density must be treated as independent of frequency, as it is the temperature of the bath that fluctuates. Actually, the differential Eq.~(\ref{eq:EQUADIFF2}), which we will derive later, indicates that. Consequently, the energy spectral density is expressed in terms of the new admittance. By definition of the admittance with $\xi$ and $F$ as fluctuating variables:

\begin{equation}
  \label{eq:SKSIKSISFF}
  S_{\xi\xi}(\omega)=S_{FF}(\omega)\frac{|Y_{s}(\omega)|^{2}}{\omega^{2}} .
\end{equation}

The integration over all the positive frequencies yields to the expected formula~(\ref{eq:GARDENKSI}):

\begin{equation}
  \label{eq:GARDENKSI2}
  \overline{\left(\delta\xi\right)^{2}}=\frac{1}{2\pi}\displaystyle \int_{0}^{+\infty} d\omega S_{FF}(\omega)\left|Y_{s}(\omega)\right|^{2}\omega^{-2} . \nonumber
\end{equation} 

There is no interest to take the real part of $S_{\xi\xi}$ which is a real number, because it comes from the autocorrelation function of $\delta\xi(t)$ for the specific time drift $\tau=0$ for which the function is maximum. This is the classical approach \cite{KITTEL1,REIF,MACCOMBIE}.

Let us apply it for thermal variables where Eq.~(\ref{eq:SKSIKSISFF}) above transforms to:

\begin{equation}
  \label{eq:SEESTT}
  S_{EE}(\omega)=\frac{S_{TT}(0)}{T^{4}} \frac{|Y_{s}(\omega)|^{2}}{\omega^{2}}.
\end{equation}

Here, $S_{TT}(0)=4k_{B}T^{2}/K$ is chosen by assumption of the series-system, such as the $\omega$-dependency of the energy spectral density comes only from the term $|Y_{s}(\omega)|^{2}/\omega^{2}$. This latter has to be evaluated for thermal bath temperature fluctuations. For that, let us consider the thermal bath which by definition has a specific heat much larger than that of the dissipative system ($C_{b}>>C$). Such as depicted in Fig~(1b), the fluctuating power of the bath can be decomposed in two components:

\begin{equation}
  \label{eq:POWERB}
  \delta P_{b}(t)=C_{b}\frac{\delta T_{b}}{dt}+C\frac{\delta T}{dt}  ,
\end{equation}
where the subscript $b$ is associated to the bath. Hence the small part of the power that induces fluctuations in the system is just the second term of the right-hand-side of this equality, $\Delta P_{b}(t)=C\delta T/dt$. Since in a series-system there is no fortuitous force at the level of the system itself, other than that coming from the bath, the differential equation~(\ref{eq:EQUADIFF}) becomes:

\begin{equation}
  \label{eq:EQUADIFF2}
  C\frac{\delta T}{dt}+K\delta T=K\delta T_{b}  , 
\end{equation}
where now the temperature bath fluctuations are considered.

Performing the Fourier transformation of the previous equation and of $\Delta P_{b}(t)$ yields to both equations:

\begin{subequations}
\begin{equation}
  \label{eq:DELTAPB}
  \Delta P_{b}(\omega)=i\omega C\delta T(\omega) ,
\end{equation}
\begin{equation}
  \label{eq:DELTAT}
  \delta T(\omega)=\frac{\delta T_{b}(\omega)}{1+i\omega\tau_{th}}                       ,
\end{equation}
\end{subequations}
where $\tau_{th} =C/K$ is the thermal relaxation time.

The admittance is the bath's admittance in a series-system:

\begin{equation}
  \label{eq:ADMITB}
  Y_{s}(\omega)=-T^{2}\frac{\Delta P_{b}(\omega)}{\delta T_{b}(\omega)}=-i\omega T^{2}\left(\frac{C}{1+i\omega\tau_{th}}\right)  . 
\end{equation}

Including the modulus of this admittance in Eq.~(\ref{eq:SEESTT}) gives directly the energy spectral density:

 \begin{equation}
  \label{eq:SEE}
  S_{EE}(\omega)=S_{TT}(0)\frac{C^{2}}{1+(\omega\tau_{th})^{2}}.
\end{equation}

Integrating this result over the positive frequencies, with $S_{TT}(0)=\frac{4k_{B}T^{2}}{K}$, gives the desired formula~(\ref{eq:EFLUC}):

\begin{equation}
  \label{eq:EFLUCBIS}
  \overline{\left(\delta E\right)^{2}}=k_{B}T^{2}C . \nonumber
\end{equation} \\

From those results, two remarks can be made:

\begin{itemize}
    \item Firstly, from Eq.~(\ref{eq:ADMITB}), the frequency dependent specific heat for the series-system is (see also \cite{GARDEN2, GARDEN3}):
\begin{equation}
  \label{eq:CSERIE}
  C_{s}(\omega)=\frac{C}{1+i\omega\tau_{th}}  .
\end{equation}
This is the part of the specific heat of the dissipative system that responds to the temperature fluctuations of the thermal bath. At high frequency on the spectrum, the system is completely insulated from bath temperature fluctuations and there is no temperature fluctuations of the system. At low frequency on the spectrum, fluctuations of the bath temperature drive the fluctuations of the temperature of the system, and $C_{s}(\omega)$ tends to the equilibrium specific heat $C$.  
    \item Secondly, it could be noticed that Eq.~(\ref{eq:SEE}) can also be written as follows:
 \begin{equation}
  \label{eq:SEE2}
  S_{EE}(\omega)=S_{TT}(\omega)C^{2},
\end{equation}
where the temperature spectral density is that of a parallel-system while the energy spectral density is that of a series-system. It provides a simple relation between fluctuating energy of a series-system and fluctuating temperature of a parallel-system implying only $C$, the equilibrium specific heat of the system considered.  \end{itemize}

To conclude this section, utilizing the following relation based on Eq.~(\ref{eq:ADMITC}):
\begin{equation}
  \label{eq:LINKYC}
  |Y_{s}(\omega)|^{2}=\omega^{2}T^{4}|C_{s}(\omega)|^{2} ,
\end{equation}
it is rather straightforward to transform Eq.~(\ref{eq:GARDENKSI2}) leading after integration to the expected formula~(\ref{eq:C1}):

\begin{equation}
  \label{eq:C1BIS}
  \overline{\left(\delta E\right)^{2}}=\frac{1}{2\pi}\displaystyle \int_{0}^{+\infty} d\omega S_{TT}(\omega)\left|C_{s}(\omega)\right|^{2} . \nonumber
\end{equation}

\subsection*{Entropy production for thermal variables fluctuations}

The equation~(\ref{eq:ENTROPYPROD2}) means that, on average over long times, or over all the possibilities of fluctuations of two conjugate thermodynamic variables, the entropy of the system decreases by a factor $-k_{B}/2$ per degree of freedom \cite{PRIGO,DEGROOT}.
In response to this reduction, the system actively generates entropy through the process of dissipation.  This relationship between fluctuations and dissipation ensures the system's proximity to its equilibrium state. It is essential to note that fluctuations exhibit correlations on a very brief time scale, where $\psi_{\xi F}(\tau) $ represents a rapidly decreasing function of the time drift $\tau$. Conversely, entropy production takes place as an average over time, extending beyond the relaxation time of the system, and thus well beyond the time scale of correlations. Throughout the relaxation process, the system's entropy production rate, denoted as entropy production because it is always positive, is observed as: 

\begin{equation}
  \label{eq:ENTPROD1}
  \sigma_{i}=\frac{d_{i}S}{dt}=\frac{d(\Delta_{i}S)}{dt}=-\frac{1}{2}\delta F\frac{d(\delta\xi)}{dt}>0 .
\end{equation}

This is simply the instantaneous rate of generation of entropy due to the dissipation process as a consequence of fluctuations of two conjugate variables evolving along time. The minus sign ensures the system returning toward equilibrium with a negative flux for positive force fluctuation and \textit{vice et versa}. In the case of thermal variables, the process of entropy generation is equivalent to the process of dissipation across heat relaxation (on average) to the thermal bath. For conjugate variables involved in work-exchanges, entropy production is also equivalent to dissipation process with a transfer of work to heat inside the system with finally relaxation of heat to the bath (cf. appendix A for electrical variables). Since entropy production involves the intensive variable and the flux of the extensive variable, the cross-correlation function between "force" and "flux" is now employed:

\begin{equation}
  \label{eq:INTERFORCEFLUX1}
  \psi_{\dot{\xi}F}(\tau)=\lim_{{t' \to +\infty}}\frac{1}{t'} \displaystyle \int_{0}^{t'} dt \delta \dot{\xi}(t)\delta F(t-\tau) .
\end{equation}

The same method than before provides for $\tau=0$:

\begin{equation}
  \label{eq:INTERFORCEFLUX2}
  \overline{\sigma_{i}}=-\frac{\psi_{\dot{\xi}F}(0)}{2}=-\frac{1}{4\pi}\displaystyle \int_{0}^{+\infty}d\omega \Re\{S^+_{\dot\xi F}(\omega)\} ,
\end{equation}
where, by means of the hermitic symmetry, $\Re\{S^+_{\dot\xi F}(\omega)\}$, is the real part of the flux and force cross-spectral density, restricted on positive frequencies. Owing to the definition~(\ref{eq:ADMIT}) of the admittance, the real part of $S_{\dot\xi F}=\delta\dot\xi(\omega)\delta F^*(\omega)$ is transformed leading to the desired formula~(\ref{eq:ENTROP1}):

\begin{equation}
  \label{eq:ENTROP1BIS}
  \overline{\sigma_{i}}=-\frac{1}{4\pi}\displaystyle \int_{0}^{+\infty} d\omega S_{\dot{\xi}\dot{\xi}}(\omega)\Re\{Z_{p}(\omega)\} , \nonumber
\end{equation}
with $Z_{p}(\omega)=1/Y_p{(\omega)}$. The average entropy production is in connection with the real part of the generalized impedance, \textit{i.e.} the dissipative part. Owing to the relation~(\ref{eq:ADMITC}) between the admittance and the complex specific heat and Eq.~(\ref{eq:ACCALO}), it is straightforward to derive the expected formula~(\ref{eq:ENTROP2}):

\begin{equation}
  \label{eq:ENTROP2BIS}
  \overline{\sigma_{i}}=\frac{1}{4\pi T^{2}}\displaystyle \int_{0}^{+\infty} d\omega S_{TT}(\omega)\omega C"_{p}(\omega) . \nonumber
\end{equation}

This confirms that the frequency dependent specific heat is a susceptibility since its imaginary part is involved in the production of entropy, or equivalently the dissipation process. An application of this last formula for the thermal process governed by Eq.~(\ref{eq:EQUADIFF}) with $C''_{p}(\omega)=K/\omega$ and with Eq.~(\ref{eq:STT}) of the frequency dependent temperature spectral density, reduces, after integration over the frequency, to the very simple result:

\begin{equation}
  \label{eq:ENTPRODMEAN}
  \overline{\sigma_{i}}=\frac{k_{B}}{2\tau_{th}} .
\end{equation}

The mean dissipated power involved is $\overline{P_{i}}=T\overline{\sigma_{i}}$. However, as we have considered on averaging times much longer than the relaxation time of the process $\tau_{th}$, there is no production of entropy anymore for $t>\tau_{th}$. This means that the mean energy involved is:

\begin{equation}
  \label{eq:EMEAN1}
  \overline{E_{i}}\sim \overline{P_{i}}\tau_{th}=\frac{k_{B}T}{2} .
\end{equation}

We recover the equipartition theorem with average energy of $k_{B}T/2$ per degree of freedom. We can conclude that the formulas~(\ref{eq:ENTROP1}) and~(\ref{eq:ENTROP2}) (or Eq.~(\ref{eq:ENTROP1BIS})) above are particular expressions of the equipartition theorem with equal weight in energy repartition at thermodynamic equilibrium.

Finally, in focusing on a specific Fourier's component on the spectrum at a given frequency $\omega_{0}=2\pi/T_{e}$, the net positive amount of entropy generated over the period $T_{e}$ can be calculated:

\begin{equation}
  \label{eq:EMEAN2}
  \left.\Delta_{i}S\right|_{\omega=\omega_{0}}=\left.\overline{\sigma_{i}}\right|_{\omega_{0}}\times T_{e}    .
\end{equation}

Excluding the integrand from the integral in formula~(\ref{eq:ENTROP2}) just above for the particular value $\omega=\omega_{0}$ (per unit of frequency $\Delta f$) gives:

\begin{equation}
  \label{eq:PRODOMEGAO}
  \left.\overline{\sigma_{i}}\right|_{\omega_{0}}=\frac{1}{2T^{2}}S_{TT}(\omega_{0})\omega_{0}C"_{p}(\omega_{0})      ,
\end{equation}
which yields to the following result:

\begin{equation}
  \label{eq:ENTROPOMEGA}
  \left.\Delta_{i}S\right|_{\omega=\omega_{0}}=\pi\frac{S_{TT}(\omega_{0})}{T^{2}}C"_{p}(\omega_{0})    .
\end{equation}

Here again, this is the analogue of a known relation obtained in modulated calorimetry \cite{GARDEN1,GARDEN2,GARDEN3,GARDEN4}, replacing the temperature spectral density by the square modulus of the oscillating sample's temperature:

\begin{equation}
  \label{eq:ENTROPOMEGA}
  \left.\Delta_{i}S\right|_{\omega=\omega_{0}}=\pi \left(\frac{\delta T{ac}}{T}\right)^{2} C"_{p}(\omega_{0})    .
\end{equation}

 It was proven in Ref.~\cite{GARDEN2} that this later formula applies to very different thermal processes, such as for example the heat propagation process investigated in appendix C.

\section{Discussion}
\label{sec:Discuss}

What is the link between the fluctuations of the state variables of a thermodynamic system at equilibrium and the macroscopic response of these same variables when the system is perturbed away from equilibrium by an external force? This connection is characterized by the process of dissipation. For a system to be in a state of thermodynamic equilibrium, and that it remains indefinitely in this state, the presence of dissipative processes is fundamental. The internal redistribution of heat within the system (energy equipartition) counteracts the effect of fluctuations maintaining a constant temperature. Despite the fact that the underlying process is the same, the order of magnitude of the power associated with dissipation during fluctuations is negligible compared to the power involved in the macroscopic response of the system to an external disturbance. It is on the order of $k_{B}T$ per hertz of bandwidth and per degree of freedom at low frequency (approximately $4\times10^{-21}$W at room temperature; see also appendix A). 

At the macroscopic level, the dissipation process is clearly expressed through the linear response theory \cite{POTTIER}. This theory shows that the dissipation of heat when the system is perturbed by an external field is linked to the imaginary part of the generalized susceptibility. If we denote $a(t)$ as the external field applied to the system's Hamiltonian, such that the harmonic perturbation is represented by $\Re\{a\exp -i\omega t \} $, then the average power dissipated in the system at angular frequency $\omega_{0}$ takes a simple form \cite{POTTIER}: 

\begin{equation}
  \label{eq:DISSIP1}
  \overline{P_{i}}=\frac{1}{2}a^{2}\omega_{0}\Im\{ \chi(\omega_{0}) \}   ,
\end{equation}
where $\Im\{ \chi(\omega)  \}$ is the imaginary part of the generalized susceptibility. However, as the Hamiltonian at equilibrium is perturbed by a term containing an externally coupled field to the fluctuating variable, we consider solely the exchange of work with the system in equilibrium. The macroscopic thermal variables "heat" and "temperature", on the other hand, are involved in heat exchange between a macroscopic system coupled to the thermal bath. As these two variables are not Hamiltonian by definition, linear response theory cannot apply \cite{POTTIER,KUBO1}. Nevertheless, Kubo's approach, based on the expression of entropy production associated with an equivalent Hamiltonian of perturbation, allows obtaining, for example, the expression of thermal conductivity from equilibrium correlation functions of Fourier transforms of heat fluxes (Green-Kubo formulas, which also applies to all transport processes) \cite{POTTIER,KUBO1,KUBO2}. However, when comparing formula~(\ref{eq:DISSIP1}) to $T$ times the term $\left.\overline{\sigma_{i}}\right|_{\omega_{0}}$ in Eq.~(\ref{eq:PRODOMEGAO}), a striking similarity is observed by setting the field $a(t) = 1/T(t)$ and using relation $\chi(\omega)=-T^{2}C(\omega)$ linking the generalized susceptibility and frequency-dependent specific heat. This suggests that a linear response theory may apply for thermal variables. Nielsen and Dyre applied the linear response theory to thermal variables perturbing a statistical system obeying a set of master equations \cite{NIELSEN}. These master equations govern the temporal variations of the probabilities of each state and their transitions from one state to another. A fluctuation-dissipation relation was thus derived for the frequency-dependent specific heat (Eq.~(26) in Ref.~\cite{NIELSEN}):

\begin{align}
  \label{eq:EQUANIELSEN}
  C(i\omega) = &\frac{\left\langle (\Delta H)^{2}\right\rangle _{eq}}{k_{B}T^{2}} \nonumber \\
              &-\frac{i\omega}{k_{B}T^{2}}\int_{0}^{+\infty}dt\left\langle \Delta H(0)\Delta H(t)\right\rangle _{eq}e^{-i\omega t} \nonumber   .
\end{align}

This represents the frequency response to the perturbation of the system's energy. The first term on the right-hand side of the equation corresponds to the equilibrium specific heat of a canonical ensemble \cite{GIBBS}. The second term on the right-hand side of the equation is linked to the autocorrelation function of energy (at constant volume) or enthalpy if pressure is constant. Therefore, this approach is analogous in terms of linear response with the approach employed in the second part of the paper for a series-system, where the temperature of bath fluctuates. Thus, the expression above should be similar to Eq.~(\ref{eq:C1}), even though it is not obvious at first glance.

When considering fluctuations in the temperature of the bath, the question of fluctuations in the temperature of the system itself seems no longer relevant. There have been numerous discussions on the physical reality of temperature fluctuations in a system. However, with sensitive and stable instruments, these fluctuations are measured and the formula~(\ref{eq:TFLUC}) is approved \cite{LIPA1}. On a theoretical point of view, the following references are instructive concerning this type of debate, particularly the dispute between C. Kittel and B.B. Mandelbrot \cite{KITTEL2,MANDELBROT,KITTEL3} and the disagreement between Kittel and McFee on the same point \cite{KITTEL4,MCFEE1} based on references \cite{MCFEE2,MAZO}. 

It is not our objective here to enter into such discussion. We will simply remark that energy fluctuations in a system are obtained through a series-system in which the bath temperature fluctuates and where energy fluctuations are obtained from canonical Gibbs's distribution \cite{GIBBS}. On the other hand, temperature fluctuations are obtained through a parallel-system where a fictitious noisy power is injected directly at the system level with a bath of constant temperature. This is a very general aspect that indeed pertains to all thermodynamic variables. When seeking fluctuations in an extensive thermodynamic variable, one must consider a 'series' situation with canonical distributions, whereas when seeking fluctuations in the conjugate intensive variable, one must consider a parallel system where the composing elements are in parallel so that the flow of the extensive variable can pass through to the bath. In the latter case, entropy production is considered by means of cross-correlation functions.

In the dispute from previous references, another crucial question has been addressed: what is the minimum size for defining the temperature of such a system \cite{BERTSCH, FESHBACH, GYFTOPOULOS}? This question is of paramount importance, and numerous recent experimental studies approach it using sophisticated micro-devices \cite{COLLIN, PEKOLA1,PEKOLA2,BOURGEOIS1,BOURGEOIS2}. Noteworthy are the experimental observations at very low temperatures of the spectral density of a single phononic mode coupled to a bath, posing a multitude of fundamental questions \cite{COLLIN}. In particular, how can we define temperature fluctuations for one single phononic mode \cite{COLLIN}? Fine analysis of electronic temperature fluctuations in mesoscopic tunnel junction at low temperatures has allowed to discriminate between electron-phonon coupling and electron-photon radiative regime as a function of bath temperature, and to study electron-temperature fluctuations under non-equilibrium effective temperature conditions \cite{PEKOLA2}. Similarly, one must consider the significance of temperature in the context of a one-dimensional phonon waveguide that connects two thermal reservoirs with well-defined temperatures \cite{BOURGEOIS1}. On a local level, defining temperature becomes a challenge in small structures at low temperatures, particularly when the mean free path of phonons exceed the scale of the nano-structure, indicating the ballistic regime of phononic transport \cite{BOURGEOIS2}. Furthermore, even at these small scales, the determination of a material's thermal conductivity is intricately linked to the system's dimensions, as the mean free path is temperature and geometry-dependent under such conditions  \cite{BOURGEOIS2}. 

All these questions are only just beginning to be addressed experimentally. They closely involve the concepts of temperature, energy transport, and entropy in small systems. Our approach may provide answers through spectral analysis of the various noise sources experimentally measured in these systems. Indeed, given the existence of various thermal processes, the general formulas presented here, expressed in terms of frequency-dependent specific heat, can prove to be useful. Extended measurements of temperature or energy spectral densities may reveal unexpected behaviors. For example, in the case of Cattaneo-Vernotte equation of propagation of heat, the spectral densities are given by Eqs.~(\ref{eq:VERNOTTESTT}) and~(\ref{eq:VERNOTTESPP}) in appendix C. We can also cite, in a non-exhaustive manner, the following works on general physical phenomena such as, the violation of the fluctuation-dissipation theorem in heat transport of mesoscopic constriction between two equilibrium reservoirs from the spectral density of fluctuations of the energy flux \cite{PEKOLA3}, the spectral analysis of temperature fluctuations in two thermally coupled high resolution magnetic-salt-thermometers \cite{LIPA2}, the observation of increase temperature fluctuations during DNA thermal denaturation by means of the power spectral density of voltage fluctuations of a platinum differential calorimeter \cite{NAGAPRIYA}, the frequency spectrum of the amplitude response function with a differential calorimeter on biological macromolecular system such as phospholipid phase transitions \cite{FREIRE}. This latter work describes merely a multi-frequency specific heat calorimeter. Although more difficult to carry out (since fluctuations are generally small), direct spectral analysis of temperature noise applies to all frequencies. Therefore, it gives access to every transient thermal events, and not to relaxation processes occurring at one or few particular frequencies. The spectral analysis of temperature fluctuations must reveal all "accidental events", whatever they are, in the numerous possible paths followed by heat-carriers in a device (thermal contacts, mean free paths, barriers,...), or during a phase transition induced by external parameters (magnetic field, electric field, pressure disturbances,...).

To conclude, starting from the form of equation~(\ref{eq:EQUADIFF}), we applied the Langevin-Einstein approach, which invokes a fictitious fluctuating force to deduce its spectral density under certain assumptions. This equation applies to a so-called parallel-system, and thus temperature fluctuations holds. When observing the presence of actual temperature noise on experimental devices such as bolometers, different groups mentioned an incidental fluctuating force precisely represented by the term $\delta P(t)$ in Eq.~(\ref{eq:EQUADIFF}) \cite{MACCOMBIE,MILATZ,MACDONALD}. This allowed them to recover Eq.~(\ref{eq:STT}) for the temperature spectral density. To achieve this, they used \textit{de facto} Eq.~(\ref{eq:TFLUC}) based on the energy equipartition theorem. On the contrary, our approach is based on the cross-correlation function of energy and temperature in order to obtain the fluctuation formulas~(\ref{eq:EFLUC}),~(\ref{eq:TFLUC}) and~(\ref{eq:PFLUC}) from the spectral densities of energy, temperature and power. For power and temperature spectral densities, we specified that the system must be of a parallel-type. This means that, even at very low frequencies, the fluctuating extensive variable can "flow" from the system to the bath. For thermal variables specifically, this flux is directly a heat-flux and thus a power. For electrical variables, this flux is the electric current (whose spectral density is proportional to power). This allowed us to recover all classical fluctuation formulas. For energy spectral density, a series-type system must be considered, with fluctuations of the intensive variable of the bath driving the system's intensive variable, and thus inducing extensive variable fluctuation like it is always allowed for canonical ensembles. From cross-correlation functions, which give an idea of how are correlated two different random variables, we found fluctuation-dissipation relations for these types of parallel systems that obey a certain type of first-order differential equation like Eq.~(\ref{eq:EQUADIFF}) or Eq.~(\ref{eq:EQUADIFFRC1}). Applying these relations to thermal variables, we found several new fluctuation-dissipation relations involving frequency-dependent specific heat. In particular, formula~(\ref{eq:C3}) seems very general as it relates the Boltzmann constant to the temperature spectral density and the real part of the complex specific heat. Finally, through this approach, we obtained fluctuation-dissipation relations for entropy production in a general case, or by introducing complex specific heat for thermal variables. In particular, formula~(\ref{eq:ENTROP2}) for the average entropy production rate involves the temperature spectral density and the imaginary part of the complex specific heat. It is to be compared with relation~(\ref{eq:C3}), which involves the real part of the complex specific heat. From this relation~(\ref{eq:ENTROP2}), by integrating over a specific period (\textit{i.e.}, a specific frequency $\omega_{0}$), we were able to retrieve a known formula in the field of oscillating temperature calorimetry, representing the net entropy produced per period \cite{GARDEN1,GARDEN2,GARDEN3,GARDEN4}. This once again proves that, starting from the fluctuations of a system in equilibrium, or from the perturbation of this system by a macroscopic external action, the same result is achieved. Formula~(\ref{eq:ENTROP1}) (or formula~(\ref{eq:ENTROP2}) for thermal variables) is thus the analogue of the central formula~(\ref{eq:DISSIP1}) of linear response theory. However, formula~(\ref{eq:ENTROP1}) arises from fluctuations and not from macroscopic perturbation. It is very general, providing the average entropy production rate emitted by a fluctuating system to maintain it in equilibrium. It applies to all phenomena involving a set of fluctuating conjugate variables in the thermodynamic sense described by differential equations of the type like in Eqs.~(\ref{eq:EQUADIFF}) or~(\ref{eq:EQUADIFFRC1}). We have also applied it to the case of an electrical simple $RC$ electrical circuit with a $RC$ parallel configuration, where known results were recovered (including the two Johnson/Nyquist formulas; cf. annexe A)). Regarding thermal variables, all these different relations can apply not only to the classical case described in figure~(\ref{fig:Fig1}) but also to a wide class of thermal processes involving energy and temperature. An example is provided in appendix C, where we have obtained expressions for the power and temperature spectral densities of a system subjected to a fluctuating power obeying the Cattaneo-Vernotte propagation equation. 

\begin{acknowledgments}
The author would like to thank D. Bourgault, O. Bourgeois, E. Collin, H. Guillou, and J. Richard (Institut Néel, Grenoble, France) for a careful reading of the manuscript and judicious comments. 
\end{acknowledgments}

\appendix
\section{Application to a RC electrical system}
\label{A}

The Fig. (2a) and (2b) depict an electrical $RC$ circuit in parallel and series mode respectively. 
\begin{figure}[h]
  \centering
  \includegraphics[width=8.6 cm]{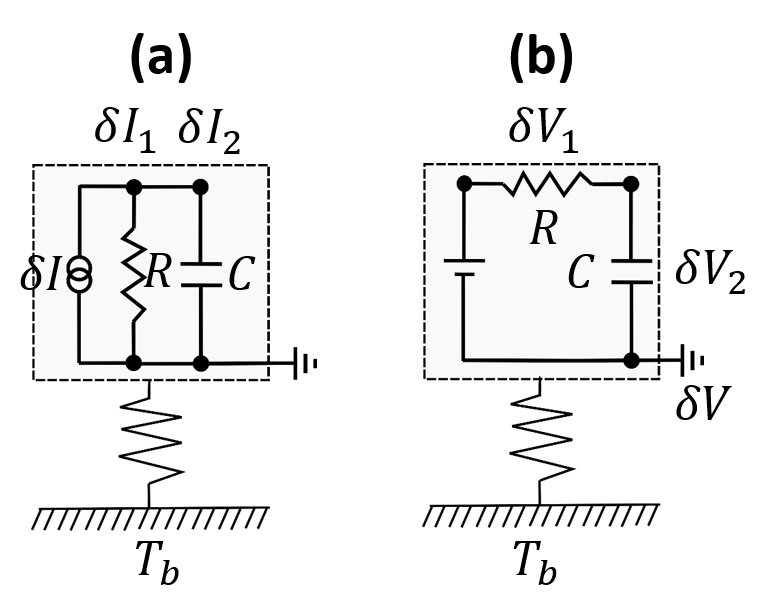}
  \caption{Picture of the dissipative system composed of pure resistor $R$ and a condenser $C$ connected to an electrical mass (electronic bath). The system is thermally coupled to a heat bath of constant temperature $T_{b}$ (a) Parallel-system. A fictitious current $\delta I$ separates in two contributions $\delta I_{1}$ and $\delta I_{2}$ in parallel in the resistor and the condenser, with one voltage fluctuations $\delta V$. (b) Series-system. A fictitious voltage $\delta V$ separates in two contributions $\delta V_{1}$ and $\delta V_{2}$ in series across the resistor and the condenser, with one current fluctuations $\delta I$ and one charge fluctuations $\delta q$.}
  \label{fig:Fig2}
\end{figure}
The thermodynamic dissipative system is composed of a pure resistor connected with a condenser, both being at temperature $T$ connected to a thermal bath, and also in connection with a charge reservoir (voltage bath). As the thermal bath is composed of a large amount of particles with respect to the dissipative system, the voltage bath is composed of a large amount of electrons with respect to the number of electrons in the system. The extensive variable is the charge $q$ of the electrons. The conjugate intensive variable is the voltage $V$ divided by temperature with a minus sign accounting for work exchange and positive dissipated work, $\delta F=-\frac{\delta V}{T}$ \cite{CALLEN4}. In Fig. (2a), the random voltage due to incessant exchanges of electrons with the electronic bath obeys to the following differential equation:

\begin{equation}
  \label{eq:EQUADIFFRC1}
  C\frac{d(\delta V)}{dt}+ \frac{\delta V}{R}=\delta I .
\end{equation}

The correspondance is immediate with Eq.~(\ref{eq:EQUADIFF}) where the fictitious force being the charge flux and not the heat-carriers flux. The thermodynamic admittance is:

\begin{equation}
  \label{eq:ADMITTRC}
  Y_{p}(\omega)=-T\frac{\delta I(\omega)}{\delta V(\omega)}=-\frac{T}{R}-i\omega CT .
\end{equation}

With $S_{II}$ the spectral density of the fictitious force being constant by assumption, the formula~(\ref{eq:GARDENKB}) yields, after integration over frequency of the remaining contribution $\frac{\Im\{Y_{p}(\omega)\}}{\omega|Y_{p}(\omega)|^{2}}$, to:

\begin{equation}
  \label{eq:SII}
  S_{II}=\frac{4k_{B}T}{R},
\end{equation}
which by a new integration over frequency leads to the analogue of formula~(\ref{eq:PFLUC}) for the electrical circuit:

\begin{equation}
  \label{eq:IFLUC}
  \overline{\left(\delta I\right)^{2}}=\frac{4k_{B}T\Delta f}{R} .
\end{equation}

With this expression of the current spectral density, the integrand in formula~(\ref{eq:GARDENF}) is calculated leading to the voltage spectral density:

\begin{equation}
  \label{eq:SVV}
  S_{VV}(\omega)=\frac{4k_{B}TR}{1+(\omega RC)^{2}} .
\end{equation}

The spectral density of voltage fluctuations is frequency dependent. Remarking that the real part of the electrical impedance is $\Re\{Z_{e}(\omega)\}=-T\Re\{Z_{p}(\omega)\}=R/(1+(\omega RC)^{2})$, then the mean square voltage fluctuations can be written:

\begin{equation}
  \label{eq:VFLUC1}
  \overline{\left(\delta V\right)^{2}}=\frac{2k_{B}T}{\pi}\displaystyle \int_{0}^{+\infty} d\omega\Re\{Z_{e}(\omega)\} .
\end{equation}

This is the generalization of the Johnson/Nyquist formula for a circuit with a real part of the impedance which depends on frequency, which is the case for the circuit in Fig.~(2a). At low frequency, on a bandwidth where the condenser does not play any role this leads directly to:

\begin{equation}
  \label{eq:VFLUC2}
  \overline{\left(\delta V\right)^{2}}=4k_{B}TR\Delta f ,
\end{equation}
which is certainly the most famous of the two formulas in the articles of Johnson and Nyquist \cite{JOHNSON,NYQUIST}. However, integrating the real part of $Z_{e}(\omega)$ over frequencies gives for the mean square value of the voltage fluctuations:

\begin{equation}
  \label{eq:VFLUC3}
  \overline{\left(\delta V\right)^{2}}=\frac{k_{B}T}{C}  .
\end{equation} \\

Continuing our procedure, let us flip to the series-circuit in Fig.~(2b), where the electronic bath imposes voltage fluctuations. The resulting random current in the circuit obeys then to the following equation:

\begin{equation}
  \label{eq:EQUADIFFRC2}
  R\delta I+ \displaystyle \int dt\frac{\delta I}{C}=\delta V  ,
\end{equation}
from which the thermodynamic impedance is:

\begin{equation}
  \label{eq:ZSERIES}
  Z_{s}(\omega)=-\frac{1}{T}\frac{\delta V}{\delta I}=-\frac{R}{T}-\frac{1}{i\omega CT}  .
\end{equation}

The formula~(\ref{eq:GARDENKSI}) with the use of Eq.~(\ref{eq:SVV}) at $\omega=0$ gives after integration of the remaining integrand $\left|Y_{s}(\omega)\right|^{2}\omega^{-2}$, the expression of the mean square value of the charge fluctuations:

\begin{equation}
  \label{eq:QFLUC}
  \overline{\left(\delta q\right)^{2}}=k_{B}TC  .
\end{equation}
We used Eq.~(\ref{eq:SVV}) at $\omega=0$ by assumption of a fictitious constant random voltage in the series-system obeying Eq.~(\ref{eq:EQUADIFFRC2}) (See also Fig.~(2b)), where the Johnson/Nyquist formula ~(\ref{eq:VFLUC2}) holds.

The application of formula~(\ref{eq:ENTROP1}) to the parallel-circuit in Fig.~(2a), with Eq.~(\ref{eq:SII}) gives the same results than for thermal variables for the average entropy production over long time (cf. Eq.~(\ref{eq:ENTPRODMEAN})) :

\begin{equation}
  \label{eq:ENTPRODMEAN2}
  \overline{\sigma_{i}}=\frac{k_{B}}{2\tau_{e}} .
\end{equation}

$\tau_{e}=RC$ is here the electronic relaxation time, generally well smaller than thermal relaxation time. This is why the formula~(\ref{eq:VFLUC2}) is valid over a broad range of experimental frequencies. Over this frequency range, for a pure resistor, the integrand $S_{II}\Re\{Z_{p}(\omega)\}$ is constant, such that the calculation of formula~(\ref{eq:ENTROP1}) yields to the interesting result for the mean entropy rate:

\begin{equation}
  \label{eq:ENTPRODMEAN3}
  \overline{\sigma_{i}}=4k_{B}\Delta f ,
\end{equation}
independent of temperature, resistance and capacitance. 

On the other hand, the dissipated power depends on temperature via $\overline{P_{i}}=T\overline{\sigma_{i}}=4k_{B}T\Delta f$. On this frequency range, the power is dissipated as a transfer of a small electrical work to a small amount of heat relaxing to the thermal bath. As already said this entropy production process allows the system to remain in equilibrium to the temperature $T_{b}$ of the bath. For a system in a non-equilibrium state, consisting of two resistors electrically connected, but thermally coupled to two baths at different temperatures, a respective exchange of small electrical works occurs between the resistors as well as a permanent heat-flux between them \cite{CILIBERTO1}. Conducting such highly sensitive noise experiments under these seemingly simple conditions unveils a richness of new results, providing new insights from a fundamental thermodynamic point of view at the microscopic scale \cite{CILIBERTO1}. Under these circumstances, it is possible to obtain the measurement of a non-equilibrium heat capacity as the linear energy response of the two-resistors-system under non-equilibrium stationary conditions when the temperature of one of them is changed \cite{CILIBERTO2}. In other types of non-equilibrium situations, it is rather usual that due to big dissipated power in the electronic circuit (by means of a considerable current crossing the circuit for example), the electronic system has an electron's temperature superior than that of the thermal bath (generally phonon's bath). This is rather usual at low temperature because the thermal coupling between electrons and their surrounding is weak. The process of dissipation heats the electron's temperature \cite{ROUKES,PEKOLA2}. In this case, the production of entropy inside the system allows the definition of an effective temperature \cite{GARDEN5}. 

Eventually, to be complete in the investigation of this electrical case, it could be of interest to compare the general formula~(\ref{eq:GARDENKB}) with the original formula of the mean square current fluctuation derived by Johnson and Nyquist in their famous genuine papers \cite{JOHNSON,NYQUIST}, that we write here for a sake of comparison:

\begin{align}
  \label{eq:JOHNSONNYQUIST}
  \overline{\left(\delta I\right)^{2}} = \frac{2k_{B}T}{\pi}\displaystyle \int_{0}^{+\infty} d\omega R(\omega)\left|Y(\omega)\right|^{2} ,
\end{align} 
where $R(\omega)$ in the papers is $\Re\{Z(\omega)\}$. A careful reading of the papers shows that the admittance is that of a parallel circuit, so that the formula~(\ref{eq:GARDENKB}) is useable. We write here formula~(\ref{eq:GARDENKB}) in terms of impedance for a sake of comparison, and with $S_{II}$ as the spectral density of the flux of the extensive variable:

\begin{equation}
  \label{eq:GARDENKBJOHNSON}
  \frac{1}{2\pi}\displaystyle \int_{0}^{+\infty} d\omega S_{II}(\omega)\frac{\Im\{Z(\omega)\}}{\omega}=-k_{B} .
\end{equation}

The Nyquist/Johnson formula involves the real part of the impedance while the formula~(\ref{eq:GARDENKBJOHNSON}) involves the imaginary part of the impedance. However, a second integration as to be performed from (\ref{eq:GARDENKBJOHNSON}) to obtain the mean square value of the current like in Eq.~(\ref{eq:JOHNSONNYQUIST}).  Upon the assumption $S_{II}(\omega)=S_{II}=constante$ (which is also the same result found by Nyquist and Johnson since the integrand in formula~(\ref{eq:JOHNSONNYQUIST}) does not depend on frequency), and remarking that $Z_{thermo}=-Z_{elec}/T$, then the integration over frequencies of the two types of integrals gives exactly the same result~(\ref{eq:SII}) or~(\ref{eq:IFLUC}). The formula~(\ref{eq:GARDENKB}) for the electrical case is consequently another formulation of the first of the two Johnson/Nyquist formulas~(\ref{eq:JOHNSONNYQUIST}) \cite{JOHNSON,NYQUIST}. As a final remark, the reasoning of Nyquist to obtain the formula~(\ref{eq:JOHNSONNYQUIST}) is based on the equipartition theorem. In our approach we mostly used the formula~(\ref{eq:ENTROPYPROD2}) which is demonstrated in the next appendix.

\section{Calculation of mean entropy fall due to fluctuations}
\label{B}

The entropy of a system at thermodynamic equilibrium ($S_{eq}$) is maximum. This is only true up to a first order since the intensive variables of the system and the bath are similar. On the other hand, fluctuations are the expression of a small spontaneous disturbances provoking small disequilibrium in the thermodynamic state of the system, and consequently a small decrease of the entropy around $S_{eq}$. Let an extensive state variable $x$ undergoing a fluctuation $\delta x=x-\overline{x}$. Due to this fluctuation, a series expansion of the system's entropy around equilibrium is written:

\begin{equation}
  \label{eq:SEXPENSION}
  S\sim S_{eq}+\delta S+\frac{1}{2}\delta^{2} S ,
\end{equation}
with $\delta S=0$ since $S_{eq}$ is maximum. For two conjugate variables, $\delta^{2}S=\delta x\delta y$ with $\delta y$ the fluctuation of the conjugate intensive variable associated to $x$. We have the requirement $\delta x\delta y<0$ since $S_{eq}$ is maximum. Since at equilibrium $\delta S=0$, this means that $\delta x=0$ (and $\delta y=0$). Therefore, sufficiently close to equilibrium we can always write $\delta x=-\alpha\delta y$ with $\alpha >0$, the two fluctuations vanishing at the same time at equilibrium. The entropy fall due to fluctuations is then written:

 \begin{equation}
  \label{eq:SFALL}
  \Delta_{i}S=S-S_{eq}\sim -\frac{1}{2}\alpha(\delta y)^{2}<0 .
\end{equation}

The probability density of a fluctuation $\delta y$ is:

\begin{equation}
  \label{eq:PROBABIS}
  p=Ae^{\frac{\Delta_{i}S}{k_{B}}}=Ae^{-\frac{1}{2}\alpha(\delta y)^{2}} .
\end{equation}

By assumption, this probability density is of Gaussian type. Its integration over all the possible values of the fluctuation $\delta y$ is equal to one. This provides the constant $A$:

\begin{equation}
  \label{eq:A}
  A=\frac{1}{\int_{-\infty}^{+\infty}e^{-\frac{\alpha(\delta y)^{2}}{2k_{b}}} d(\delta y)}=\sqrt{\frac{\alpha}{2\pi k_{B}}} .
\end{equation}

The expectation value of $\Delta_{i}S$ taken over all the possible fluctuations is:

\begin{align}
  \label{eq:DELTAISAVERAGE}
  \left\langle \Delta_{i}S\right\rangle=&-\frac{A}{2}\int_{-\infty}^{+\infty}\alpha(\delta y)^{2}e^{-\frac{\alpha(\delta y)^{2}}{2k_{b}}} d(\delta y)\nonumber \\
  =&-\frac{A}{2}\frac{\alpha k_{B}\sqrt{2\pi k_{B}}}{\alpha\sqrt{\alpha}} \nonumber \\
  =&-\frac{k_{B}}{2} .
\end{align}

\section{Application to Cattaneo-Vernotte equation of heat propagation}
\label{C}

\begin{figure}[h]
  \centering
  \includegraphics[width=8.6 cm]{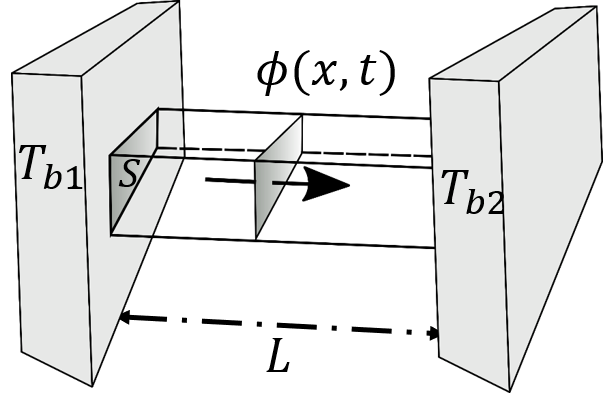}
  \caption{A rectangular slab of surface $S$ and length $L$ is in perfect thermal contact with two heat baths of temperature $T_{b1}$ and $T_{b2}$ respectively. At a length $x$ on the slab, the surface $S$ is crossed by a heat-flux $\phi(x,t)$ which is proportional to the thermal gradient $\frac{\partial T(x,t)}{\partial x}$ following the Fourier's law.}
  \label{fig:Fig3}
\end{figure}

Let us consider a slab of surface $S$ and length $L$ with its two boundaries in perfect thermal contact with two heat baths of temperature $T_{b1}$ and $T_{b2}$, such as depicted in Fig.~(\ref{fig:Fig3}). 

From an initial equilibrium state where $T_{b1}=T_{b2}$, if the temperatures of the baths are suddenly changed such that $T_{b1}>T_{b2}$, then after a certain time, a constant amount of heat flows perpendicularly to the surface of the slab (one dimensional heat flow per unit of surface). At any length $x$ this constant heat-flux is proportional to the thermal gradient following the Fourier's law: 

\begin{equation}
  \label{eq:FOURIER}
  \phi(x)=-\lambda\frac{\partial T(x)}{\partial x} .
\end{equation}

The establishment of this stationary conditions takes place after the relaxation time $\tau_{i}=L^{2}/D$, where $D$ is the thermal diffusivity of the material composing the slab. The coefficient of proportionality $\lambda$ is the thermal conductivity which is again an intrinsic property of the material constituting the slab. The Fourier's law is valid in plenty of different situations. However, it suffers from the paradoxical effect of an infinite velocity of heat propagation within the slab. In response to an instantaneous change of the thermal gradient at the coordinate $x$ corresponds an instantaneous change of the heat-flux everywhere in the slab. However, under particular physical circumstances, the heat-carriers cannot respond instantaneously to such thermal changes if it happens. For example, this happens in the case of rapid energy shots on a surface (laser pulses), or in heat propagation across the surfaces of multi-layers media after rapid temperature changes, or at low temperature when the mean free path of heat carriers becomes higher than thermal gradients imposes in the system. In such cases, the conventional Fourier's law is no longer tenable. Cattaneo and Vernotte showed independently that in order to remove this paradox of infinite velocity of propagation of heat in a body, then a supplementary term has to be added to the Fourier's law \cite{VERNOTTE,CATTANEO}:

\begin{equation}
  \label{eq:CATAVERNO}
  \phi(x,t)+\tau\frac{\partial\phi(x,t)}{\partial t}=-\lambda\frac{\partial T(x,t)}{\partial x} .
\end{equation}
with $\tau$ the relaxation time of the heat carriers (not to be confused with the diffusive relaxation time $\tau_{i}$). The heat-flux becomes a relaxing variable. A good picture could be to imagine that now the slab is a volume $V=SL$ filed with a rarefied gas, or liquid helium at low temperatures, with mean free path $l$ becoming substantial with $\partial T/\partial x>>T/l$.

However, even at equilibrium with $T_{b1}=T_{b2}$, thermal agitation of heat-carriers provokes microscopic fluctuations of heat-flux and temperature, and thus fluctuations of temperature gradient. At a position $x$ in the slab, thermal variable fluctuations obey to the following equation:

\begin{equation}
  \label{eq:VERNOTTEFLUC}
  \delta \phi(x,t)+\tau\frac{\partial\delta\phi(x,t)}{\partial t}=-\lambda\left(\frac{\partial\delta T(x,t)}{\partial x}\right)=-\lambda\delta \left(\frac{\partial T(x,t)}{\partial x}\right) ,
\end{equation}
where it has been supposed that temperature fluctuations do not induce position fluctuations $\delta x$ across thermal dilatation coefficient.

This equation is of the same form than Eq.~(\ref{eq:EQUADIFF}) or Eq.~(\ref{eq:EQUADIFFRC1}), but, this time, with $\delta \left(\frac{\partial T}{\partial x}\right)$ as the fictitious force. By taking the Fourier transform of the previous equation, a complex specific heat (per unit of length $x$) can be defined at each position $x$:

\begin{align}
  \label{eq:COMEGACATTANEO}
  C_{x}(\omega)=&\frac{\delta\phi(x,\omega)}{i\omega\delta \left(\frac{\partial T(x,\omega)}{\partial x}\right)}=\frac{\lambda}{\omega^{2}\tau-i\omega} \nonumber \\
  =&\frac{\lambda\tau}{1+(\omega\tau)^{2}}-i\frac{\lambda/\omega}{1+(\omega\tau)^{2}} .
\end{align}
It does not depend on $x$.
 
Contrary to the process following Eq.~(\ref{eq:EQUADIFF}), the form of Eq.~(\ref{eq:VERNOTTEFLUC}) suggests that the temperature gradient spectral density is constant while the heat-flux spectral density is not. This is normal because in Cattaneo-Vernotte process the heat-flux is the relaxing variable. Let us apply the formula~(\ref{eq:C3}) in excluding $S_{\frac{\partial T}{\partial x}\frac{\partial T}{\partial x}}$ from the integral in order to find:

\begin{equation}
  \label{eq:VERNOTTESTT}
  S_{\frac{\partial T}{\partial x}\frac{\partial T}{\partial x}}=\frac{2\pi k_{B}T^{2}}{\int_{0}^{+\infty}\frac{\lambda\tau}{1+(\omega\tau)^{2}}d\omega}=\frac{4 k_{B}T^{2}}{\lambda} .
\end{equation}

The mean square value of the thermal gradient fluctuation is:
\begin{equation}
  \label{eq:VERNOTTETFLUC}
  \overline{\delta\left(\partial T/\partial x\right)^{2}}=\frac{4k_{B}T^{2}\Delta f}{\lambda}.
\end{equation}

By means of Eq.~(\ref{eq:COMEGACATTANEO}) (and Eq.~(\ref{eq:ACCALO})), we obtain the frequency dependence of the heat-flux spectral density:

\begin{align}
  \label{eq:VERNOTTESPP}
  S_{\phi\phi}(\omega)=&\frac{\lambda^{2}}{1+(\omega\tau)^{2}} S_{\frac{\partial T}{\partial x}\frac{\partial T}{\partial x}} \nonumber\\
  =&\frac{4k_{B}T^{2}\lambda}{1+(\omega\tau)^{2}} ,
\end{align}
which gives after integration over positive frequency, the mean square value of the heat-flux fluctuations:

\begin{equation}
  \label{eq:VERNOTTEPFLUC}
  \overline{\left(\delta \phi\right)^{2}}=\frac{k_{B}T^{2}\lambda}{\tau}.
\end{equation}

If we compare this result with the classical formula~(\ref{eq:PFLUC}), we conclude that in the conditions of Cattaneo-Vernotte propagation of heat, the heat-flux fluctuation formula has the same form than Eq.~(\ref{eq:PFLUC}) but with a frequency bandwidth of $\Delta f = 1/4\tau$ where $\tau$ is the relaxation time of heat-carriers. With this particular example of Cattaneo-Vernotte equation, we have shown that it is possible, by means of our procedure and using frequency dependent complex specific heat, to obtain new fluctuations formulas for thermal variables. All the formulas concerning the entropy production are also valid (see reference~\cite{GARDEN2}).


\begin{thebibliography}{xx}
\bibitem{FOOTNOTE1}
when $K$ is a thermal conductance made of an insulating material, the heat-carriers are phonons, and we speak about phonon noise.
\end{thebibliography}
\end{document}